\documentclass[aps,pra,reprint,floatfix,superscriptaddress,nofootinbib,longbibliography]{revtex4-2}

\usepackage[T1]{fontenc}
\usepackage[utf8]{inputenc}
\usepackage{amsmath,amssymb,amsthm,mathtools,bm}
\usepackage{graphicx}
\usepackage{braket}
\usepackage{hyperref}
\usepackage{xcolor}

\hypersetup{
  colorlinks=true,
  linkcolor=blue,
  citecolor=blue,
  urlcolor=blue
}

\begin{document}

\title{Sector-memory obstruction to probe-level bath emergence in finite programmable qubit environments}

\author{Gaurav Sarmah}
\author{Ramakrishna Podila}
\email{rpodila@g.clemson.edu}
\affiliation{Department of Physics and Astronomy, Clemson University, Clemson, SC 29634, USA}

\begin{abstract}
Finite quantum environments can relax local probes without necessarily acting as
canonical baths.  We study this distinction using a probe qubit coupled to a
finite programmable bath of \(N\) qubits with an excitation-number-conserving
dynamics.  The conserved charge partitions the Hilbert space into sectors, and
we use the sector-resolved late-time probe population \(p_e^{(q)}\), the
sector-memory variance \(M_N\), and a global Gibbs-fit error
\(\Delta_G^{\rm global}\) as operational diagnostics of probe-level bath
emergence.  Exact simulations initialized with Haar-random pure states over each
complete fixed-charge sector recover a sector-dependent population close to
the maximally mixed-sector benchmark \(p_e^{(q)}=q/(N+1)\), giving a nonzero
Gibbs obstruction.  We then construct hardware-compatible
charge-preserving Floquet circuits based on \(R_z\) phases and \(XX+YY\)
exchange gates, validate them in ideal and noisy Qiskit simulations, and
implement the corresponding finite-depth experiments on IBM Fez.  For
\(N=4\), the measured hardware data with no symmetry breaking (\(\epsilon=0\)) give
\(M_N\simeq 0.044\), \(\Delta_G^{\rm global}\simeq 0.558\), and charge
preservation near \(0.90\) after readout mitigation.  A paired
symmetry-breaking scan (\(\epsilon \neq 0\)), implemented by bath \(R_x(\epsilon)\) rotations,
reduces \(M_N\) and \(\Delta_G^{\rm global}\) while increasing charge leakage,
but does not erase sector ordering over the accessible circuit depths.  These
results demonstrate a finite-size, probe-level obstruction to bath emergence:
local equilibration within constrained sectors is insufficient to produce a
single sector-independent Gibbs state for the probe.
\end{abstract}

\maketitle

\section{Introduction}
\label{sec:introduction}

Thermal behavior in a closed quantum system is not imposed by an external
reservoir. It must emerge from unitary dynamics, from the structure of
many-body eigenstates, and from the restriction to local or few-body
observables. This problem connects the eigenstate thermalization hypothesis
(ETH) \cite{deutsch1991,srednicki1994,rigol2008,dalessio2016}, canonical
typicality \cite{tasaki1998,goldstein2006,popescu2006}, and rigorous results
on equilibration of subsystems
\cite{reimann2008,linden2009,short2011,gogolin2016}. In the ETH picture, a
nonintegrable many-body system can act as its own bath: individual energy
eigenstates reproduce thermal expectation values of local observables, and
generic initial states relax to values determined primarily by the energy
density rather than by microscopic details of the preparation
\cite{deutsch1991,srednicki1994,rigol2008,dalessio2016}. Related behavior has
been investigated in exact diagonalization and in experiments with ultracold
atoms, trapped ions, neutral atoms, and superconducting quantum processors
\cite{genway2012,kinoshita2006,kaufman2016,neill2016,bernien2017,zhu2022}.

Thermalization is nevertheless constrained by conservation laws. Integrable
systems and systems with extensive or emergent conserved quantities may relax
to generalized ensembles rather than to a single canonical Gibbs state
\cite{rigol2007,rigol2008,cassidy2011,vidmar2016,essler2016}. The quantum
Newton's cradle \cite{kinoshita2006}, generalized Gibbs ensembles in
one-dimensional Bose gases \cite{langen2015}, and persistent memory in
many-body-localized systems
\cite{nandkishore2015,schreiber2015,smith2016,abanin2019} demonstrate that
relaxation of selected observables does not necessarily erase all information
about the initial preparation. When a conservation law is weakly broken,
observables may first relax within an approximately conserved manifold and
only later drift toward an unconstrained thermal state
\cite{mori2016,kuwahara2016,abanin2017,else2017,mori2018,mallayya2019}. Thus,
equilibration within a symmetry sector and loss of memory across different
sectors are distinct physical requirements.

This distinction becomes especially important in quantum thermodynamics,
where an environment may be finite, dynamically evolving, and appreciably
perturbed by its interaction with the system. A small discrete environment
can equilibrate a subsystem even when the resulting reduced dynamics differ
from standard infinite-bath or Born--Markov predictions
\cite{gemmer2006finitebath}. Finite heat capacity modifies ideal-reservoir
thermodynamic bounds \cite{richens2018finitebath}, while consistent
finite-bath descriptions must generally account for changes in the
environmental state and for the buildup of system--environment correlations
\cite{rieracampeny2021finitebath}. Exact spin-star and central-spin studies
provide concrete examples: finite environments can induce relaxation while
also producing non-Markovianity, recurrences, and long-time dependence on
system size or initial conditions
\cite{breuer2004spinstar,mukhopadhyay2017central}.

Experiments with engineered environments have likewise shown that a small
controlled system can display several bath-like properties without becoming
equivalent to an ideal macroscopic reservoir. Engineered dissipative maps have
been implemented in trapped ions \cite{barreiro2011opensystem}, structured
photonic reservoirs have enabled tunable Markovian-to-non-Markovian dynamics
\cite{liu2011nonmarkovian}, and the thermalization of a single spin has been
studied while increasing the size of an engineered bosonic environment
\cite{clos2016thermalization}. These works establish that finite environments
can generate dissipation, equilibration, and approximately thermal probe
states. They also show that finite-size corrections, memory, recurrences, and
environmental back-action can remain important.

The unresolved issue is therefore not simply whether a finite environment can
relax a probe. The sharper question is whether it produces the same reduced
probe state for different allowed preparations. Suppose that a conserved
charge partitions the joint probe--environment Hilbert space into sectors
labeled by $q$. The dynamics may strongly mix states within every fixed-$q$
sector and produce an approximately stationary probe state
$\rho_S^{(q)}$. Nevertheless, the collection of states
${\rho_S^{(q)}}$ may retain a systematic dependence on $q$. In that case,
the environment equilibrates the probe conditionally within each constrained
sector, but it does not act as a single preparation-independent bath.

This cross-sector requirement is complementary to other operational tests of
quantum thermal behavior. Measures of non-Markovianity diagnose temporal
memory through information backflow or failure of divisible reduced dynamics
\cite{rivas2014,liu2011nonmarkovian}. Quantum fluctuation relations test
identities involving nonequilibrium work and forward or reverse driving
protocols \cite{campisi2011,batalhao2014fluctuations}. Quantum thermometry
infers a temperature from the state or response of a calibrated probe
\cite{correa2015thermometry}. None of these tests, by itself, requires
different conserved-sector preparations to yield one common probe state.
Indeed, each diagonal probe state may separately admit an effective
temperature even when those temperatures depend on $q$ and therefore fail to
define one bath.

Here we introduce a probe-level criterion that tests this missing consistency
condition. We consider a probe qubit coupled to an environment of $N$ qubits
under excitation-number-conserving dynamics. For each sampled charge sector
$q$, we determine the long-time or finite-depth probe excitation probability
$p_e^{(q)}$. If the environment acts as a single canonical bath, the
sector-resolved probe states must be compatible with one Gibbs state. We
quantify violations of this requirement using the sector-memory variance
$M_N$, which measures the dispersion of $p_e^{(q)}$ across sectors, and the
global Gibbs-fit error $\Delta_G^{\rm global}$, which measures the obstruction
to fitting all sector-resolved probe states with one inverse temperature.
These diagnostics can remain nonzero even after the probe has relaxed within
every sector.

Programmable quantum processors provide a direct setting in which to test this
criterion. Superconducting and neutral-atom platforms have already been used
to study ergodicity, information scrambling, localization, transport, thermal
observables, and many-body entanglement
\cite{neill2016,bernien2017,arute2019,zhu2022,perrin2025,zhang2024,chang2025deep}.
The present work asks a more local question: when a finite qubit environment
is prepared in different charge sectors, does one measured probe qubit
converge to a common Gibbs state, or does it retain sector memory?

We combine exact Hamiltonian calculations, ideal and noisy circuit
simulations, and experiments on an IBM superconducting processor. The exact
analysis establishes the sector-resolved Gibbs obstruction and clarifies the
difference between Haar-sector and probe-fixed initializations. The
hardware-compatible circuit uses charge-preserving $R_z$ and $XX+YY$ gates,
allowing the same probe-level criterion to be tested at finite circuit depth
\cite{qiskit2019}. The hardware measurements, employing standard read-out mitigation principles \cite{temme2017,endo2018,kandala2019,bravyi2021,cai2023}, show clear sector-dependent
probe populations and nonzero $M_N$ and $\Delta_G^{\rm global}$, demonstrating
that the finite environment does not act as one sector-independent Gibbs bath.
A paired symmetry-breaking protocol further shows that controlled violation of
charge conservation suppresses the sector-memory diagnostics, but does not
erase them over the accessible circuit depths.

The resulting conclusion is operational and deliberately limited. A finite
programmable environment can equilibrate a local probe within each conserved
sector while failing to erase information about which sector was initially
prepared. Bath emergence therefore requires more than relaxation within each
constrained manifold: it requires cross-preparation consistency of the
reduced probe state. The diagnostics introduced here provide a direct way to
test this requirement using only sector-resolved measurements of a single
qubit.

\section{Theory and methods}
\label{sec:theory}

\subsection{Hamiltonian, basis convention, and conserved charge}
\label{sec:model}

We consider a probe qubit, denoted by $S$, coupled to a finite bath $B$ of
$N$ qubits.  The Hilbert space is
$\mathcal{H}=\mathcal{H}_S\otimes\mathcal{H}_B$, with
$\dim\mathcal{H}_S=2$ and $\dim\mathcal{H}_B=2^N$.  We set $\hbar=1$.
The Hamiltonian used in the exact calculations is
\begin{equation}
H = H_S + H_B + V,
\label{eq:fullH}
\end{equation}
with
\begin{subequations}
\begin{align}
H_S &= \frac{\omega_S}{2}\sigma_z^{(S)},
\label{eq:HS}\\
H_B &= \sum_{j=1}^{N}\frac{\omega_j}{2}\sigma_z^{(j)}
      +H_{\rm int}^{B},
\label{eq:HB}\\
V &= \sum_{j=1}^{N}g_j
\left(
\sigma_+^{(S)}\sigma_-^{(j)}
+
\sigma_-^{(S)}\sigma_+^{(j)}
\right).
\label{eq:V}
\end{align}
\end{subequations}
Here, $\omega_S$ is the probe level-splitting parameter, $\omega_j$ is the
corresponding parameter for bath qubit $j$, $g_j$ is the probe--bath exchange
coupling, and $H_{\rm int}^{B}$ is the bath-internal Hermitian interaction, $\sigma_+$ and $\sigma_-$ are occupation-raising and lowering operators.  The superscript $(S)$ labels the probe, the
superscript $(j)$ labels bath qubit $j\in\{1,\ldots,N\}$, and
$\sigma_\alpha^{(\mu)}$, with $\alpha\in\{x,y,z\}$, denotes a Pauli operator
acting on qubit $\mu$ and as the identity on all other qubits.  The only
property of $H_{\rm int}^{B}$ required below is number conservation,
$[H_{\rm int}^{B},Q_B]=0$, where $Q_B=\sum_{j=1}^{N}|1\rangle_j\langle 1|$ .  

The computational basis is defined by
$\sigma_z|0\rangle=|0\rangle$ and
$\sigma_z|1\rangle=-|1\rangle$. The corresponding computational-basis occupation operator is
\begin{equation}
n\equiv \sigma_+\sigma_-=|1\rangle\langle 1|
=\frac{\mathbb{I}-\sigma_z}{2},
\label{eq:number_operator}
\end{equation}
where $\mathbb{I}$ is the single-qubit identity.  The total occupation, which
is the conserved charge of the symmetry-preserving model, is
\begin{equation}
Q=n_S+Q_B,
\qquad
n_S=|1\rangle_S\langle 1|,
\qquad
Q_B=\sum_{j=1}^{N}|1\rangle_j\langle 1|.
\label{eq:Q}
\end{equation}
With the interaction structure in Eq.~\eqref{eq:V} and the number-conserving
bath interaction specified above, $[H,Q]=0$.

\subsection{Sector-resolved reduced states and the Gibbs criterion}
\label{sec:obstruction}

Let $\rho(0)$ be the initial density operator of the full $(N+1)$-qubit
system.  Hamiltonian evolution gives
\begin{equation}
\rho(t)=U(t)\rho(0)U^\dagger(t),
\qquad
U(t)=e^{-iHt},
\label{eq:unitary_evolution}
\end{equation}
where $t$ is time.  Because the total system is finite and closed,
$\rho_S(t)={\rm Tr}_B[\rho(t)]$ need not converge pointwise at long times.
Here, ${\rm Tr}_B$ denotes the partial trace over the bath.  We therefore use
the infinite-time average
\begin{equation}
\bar{\rho}
=
\lim_{T\rightarrow\infty}\frac{1}{T}\int_0^T\rho(t)\,dt,
\qquad
\bar{\rho}_S={\rm Tr}_B(\bar{\rho}),
\label{eq:timeavg}
\end{equation}
where $T$ is the averaging time.  Equivalently,
\begin{equation}
\bar{\rho}=\sum_{\lambda}P_{\lambda}\rho(0)P_{\lambda},
\label{eq:diagonal_ensemble}
\end{equation}
where $\lambda$ labels the distinct eigenvalues of $H$ and $P_{\lambda}$ is
the projector onto the corresponding eigenspace.  Equation~\eqref{eq:diagonal_ensemble}
remains valid in the presence of degeneracies.

For a fixed conserved-charge sector, the initial state obeys
\begin{equation}
Q\rho^{(q)}(0)=q\rho^{(q)}(0),
\qquad
\rho^{(q)}(0)Q=q\rho^{(q)}(0),
\label{eq:sector}
\end{equation}
where $q\in\{0,1,\ldots,N+1\}$ is the total computational-basis occupation.
Because $[H,Q]=0$, the state remains in the same sector.  The late-time
probe occupation in sector $q$ is
\begin{equation}
p_e^{(q)}
\equiv
{\rm Tr}_S\!\left[\bar{\rho}_S^{(q)}n_S\right],
\label{eq:peq}
\end{equation}
where ${\rm Tr}_S$ is the trace over the probe Hilbert space and
$\bar{\rho}_S^{(q)}$ is obtained from Eqs.~\eqref{eq:timeavg} and
\eqref{eq:diagonal_ensemble} for an initial state in sector $q$. It should be noted that the symbol ``excitation'' in the circuit and data analysis refers to the computational occupation $|1\rangle$, as we evaluate the Hamming
weight of measured bit strings.  For the Hamiltonian sign convention in
Eq.~\eqref{eq:HS}, however, $|1\rangle$ is the lower-energy eigenstate when
$\omega_S>0$.  Thus, the quantity denoted by $p_e^{(q)}$ should be read
as the logical-$|1\rangle$ occupation probability rather than as a statement
about energetic ordering.  
The canonical Gibbs state of the probe Hamiltonian is
\begin{equation}
\rho_S^{\rm Gibbs}(\beta)
=
\frac{e^{-\beta H_S}}{Z_S(\beta)},
\qquad
Z_S(\beta)={\rm Tr}_S\!\left(e^{-\beta H_S}\right),
\label{eq:Gibbs}
\end{equation}
where $\beta\in\mathbb{R}$ is the inverse-temperature parameter and
$Z_S(\beta)$ is the probe partition function.  With the basis and Hamiltonian
conventions above, the Gibbs probability of measuring the probe in
$|1\rangle$ is
\begin{equation}
p_\beta
\equiv
{\rm Tr}_S\!\left[\rho_S^{\rm Gibbs}(\beta)n_S\right]
=
\frac{1}{1+e^{-\beta\omega_S}}.
\label{eq:pbeta}
\end{equation}

A single sector-independent Gibbs description would require one value of
$\beta$ such that
\begin{equation}
\bar{\rho}_S^{(q)}=\rho_S^{\rm Gibbs}(\beta)
\label{eq:noGibbs}
\end{equation}
for every sector included in the comparison.  Therefore, if two sectors
$q_1\neq q_2$ satisfy
\begin{equation}
p_e^{(q_1)}\neq p_e^{(q_2)},
\label{eq:diff}
\end{equation}
then no single canonical probe state can describe both preparations.
This statement concerns the collection of sector-resolved reduced states; it
does not exclude equilibration within each individual sector.

For exact fixed-$q$ dynamics, the probe coherence in the computational basis
vanishes identically.  The $|0\rangle_S$ component is accompanied by bath
charge $q$, whereas the $|1\rangle_S$ component is accompanied by bath charge
$q-1$.  These bath subspaces are orthogonal, and hence
\begin{equation}
\langle 0|\rho_S^{(q)}(t)|1\rangle=0
\label{eq:coherence_zero}
\end{equation}
for all $t$ when the initial state has definite total charge.  Consequently,
the population measurement is sufficient to test the Gibbs obstruction in
the symmetry-preserving experiment; no unmeasured probe coherence can restore
a common Gibbs state across sectors.

\subsection{Sector-memory and global Gibbs-fit diagnostics}
\label{sec:diagnostics}

The exact-Hamiltonian and IBM Fez quantum processing unit (QPU) analyses are compared over the interior total-charge sectors $\mathcal{Q}_N\equiv{1,2,\ldots,N}$. The boundary sectors ($q=0, N+1$) are excluded from the sector-memory diagnostics because they contain only one computational-basis configuration and therefore do not permit probe--bath exchange. In the principal exact-Hamiltonian analysis, the initial pure state is sampled from the Haar measure on the complete total-charge subspace ($\mathcal H_q$). Exact-Hamiltonian calculations with the initial state of the probe fixed as ($|0\rangle_S$) are presented in the Supporting Information (Figs. S1-S5). In the QPU implementation, the probe is initialized in ($|0\rangle_S$) and the bath is prepared in a computational-basis state containing exactly $q$ occupied qubits. A summary of initial-state conventions used in the calculations and experiments is provided in Table S1. 

For any sector-dependent quantity $f_q$, define
\begin{equation}
\langle f\rangle_q
\equiv
\frac{1}{N}\sum_{q\in\mathcal{Q}_N}f_q.
\label{eq:sector_average}
\end{equation}
The sector-memory variance is
\begin{equation}
M_N
\equiv
\operatorname{Var}_{q\in\mathcal{Q}_N}\!\left[p_e^{(q)}\right]
=
\frac{1}{N}\sum_{q\in\mathcal{Q}_N}
\left(p_e^{(q)}-\langle p_e\rangle_q\right)^2.
\label{eq:M}
\end{equation}
For the exact diagonal ensemble, this quantity may be denoted
$M_N^{(\infty)}$; for a depth-$d$ circuit it is denoted $M_N(d,\epsilon)$.
A nonzero value means that the probe retains information about the initially
prepared charge sector.

The global Gibbs-fit error is
\begin{equation}
\Delta_G^{\rm global}
\equiv
\min_{\beta\in\mathbb{R}}
\max_{q\in\mathcal{Q}_N}
\left\|
\bar{\rho}_S^{(q)}-\rho_S^{\rm Gibbs}(\beta)
\right\|_1,
\label{eq:DeltaG}
\end{equation}
where
$\|A\|_1\equiv{\rm Tr}\sqrt{A^\dagger A}$ is the trace norm of operator $A$.
For diagonal probe states,
\begin{equation}
\Delta_G^{\rm global}
=
2\min_{\beta\in\mathbb{R}}
\max_{q\in\mathcal{Q}_N}
\left|p_e^{(q)}-p_\beta\right|.
\label{eq:DeltaGdiag}
\end{equation}
Because $p_\beta$ spans the open interval $(0,1)$ as $\beta$ spans
$\mathbb{R}$, the minimax population is the midpoint of the largest and
smallest sector populations.  Thus,
\begin{equation}
\Delta_G^{\rm global}
=
p_{\max}-p_{\min},
p_{\max}\equiv\max_{q\in\mathcal{Q}_N}p_e^{(q)},
p_{\min}\equiv\min_{q\in\mathcal{Q}_N}p_e^{(q)}.
\label{eq:DeltaGrange}
\end{equation}
The standard
trace-distance normalization is
\begin{equation}
D_G^{\rm global}\equiv\frac{1}{2}\Delta_G^{\rm global}.
\label{eq:DG}
\end{equation}
Both normalizations vanish under exactly the same condition.  Moreover,
$M_N>0$ implies $\Delta_G^{\rm global}>0$, because a nonzero variance requires
at least two distinct sector populations.

For a sector distribution with normalized weights $w_q\geq0$ and
$\sum_qw_q=1$, the distribution-sensitive extension is
\begin{equation}
M_N[w]
\equiv
\sum_qw_q
\left(p_e^{(q)}-\sum_{q'}w_{q'}p_e^{(q')}\right)^2,
\label{eq:weighted_M}
\end{equation}
where $q'$ is a dummy sector index.  The unweighted diagnostic in
Eq.~\eqref{eq:M} corresponds to $w_q=1/N$ on $\mathcal{Q}_N$.

\subsection{Uniform fixed-charge benchmark}
\label{sec:benchmark}

The analytically solvable reference state is the maximally mixed state in a
fixed total-charge sector,
\begin{equation}
\rho_q^{\rm unif}
\equiv
\frac{\Pi_q}{\dim\Pi_q}
=
\frac{\Pi_q}{\binom{N+1}{q}},
\label{eq:unif}
\end{equation}
where $\Pi_q$ projects onto the subspace of the full $(N+1)$-qubit Hilbert
space containing exactly $q$ computational-basis occupations.  The probe is
in $|1\rangle$ precisely when the $N$ bath qubits contain $q-1$ occupations.
Therefore,
\begin{equation}
p_e^{(q)}
=
\frac{\binom{N}{q-1}}{\binom{N+1}{q}}
=
\frac{q}{N+1},
\label{eq:sectorlaw}
\end{equation}
and the reduced probe state is
\begin{equation}
\rho_{S,q}^{\rm unif}
=
\frac{q}{N+1}|1\rangle\langle1|
+
\left(1-\frac{q}{N+1}\right)|0\rangle\langle0|.
\label{eq:rhoq}
\end{equation}
This benchmark represents complete mixing within a fixed sector, not mixing
between different sectors.  Equivalently, if \(|\psi_q\rangle\) is sampled
from the Haar measure on the complete total-charge subspace
\(\mathcal H_q=\Pi_q\mathcal H\), then
\begin{equation}
\mathbb E_{\psi_q}
\left[|\psi_q\rangle\langle\psi_q|\right]
=
\frac{\Pi_q}{\dim\mathcal H_q}
=
\rho_q^{\rm unif}.
\label{eq:haar_sector_average}
\end{equation}
Consequently, Eq.~\eqref{eq:sectorlaw} is exact for the maximally mixed sector
state and for the full-sector Haar ensemble average.  It is not, by itself, an
exact prediction for every pure state of definite total charge or for an
initialization in which the probe is constrained to \(|0\rangle_S\).  The
benchmark yields
\begin{equation}
M_N^{(\infty)}
=
\operatorname{Var}_{q\in\mathcal{Q}_N}
\left[\frac{q}{N+1}\right]
=
\frac{N-1}{12(N+1)},
\label{eq:Mbench}
\end{equation}
and
\begin{equation}
\Delta_{G,{\rm unif}}^{\rm global}
=
\frac{N-1}{N+1}
\label{eq:Deltabench}
\end{equation}
for the interior-sector set $\mathcal{Q}_N$.  These are the finite-size
reference values used in the exact and circuit analyses.

\subsection{Exact-Hamiltonian calculation}
\label{sec:exact_methods}

The exact calculation evaluates the Hamiltonian in
Eq.~\eqref{eq:fullH} without modifying its terms.  The Hilbert space is
partitioned into eigenspaces of $Q$, the commutator $[H,Q]$ is checked
numerically, and the diagonal ensemble in Eq.~\eqref{eq:diagonal_ensemble} is
constructed within each sector.  For the principal $N=8$ calculation reported
in the main text, each initial state is drawn from the Haar measure on the
complete total-charge sector,
\begin{equation}
|\psi_{q,r}^{\rm full}\rangle
=
\sum_{a=1}^{D_q}c_{a}^{(r)}|a,q\rangle,
\qquad
D_q=\binom{N+1}{q},
\label{eq:full_sector_haar_initial}
\end{equation}
where the real and imaginary parts of the unnormalized coefficients are
independent Gaussian variables and the vector is subsequently normalized.
Here, $r$ labels the random preparation and $|a,q\rangle$ runs over all basis
states of the full probe--bath sector, including both probe occupations.  We
used 16 such states for each of four Hamiltonian-disorder realizations, giving
64 samples per sector.

To test whether the main conclusion depends on this full-sector Haar
initialization, we also performed a supplementary calculation with
\begin{equation}
|\psi_{q,r}^{(0)}\rangle
=
|0\rangle_S\otimes|\phi_{B,q}^{(r)}\rangle,
\qquad
|\phi_{B,q}^{(r)}\rangle\sim
\operatorname{Haar}(\mathcal H_{B,q}),
\label{eq:probe0_bath_haar_initial}
\end{equation}
where $\mathcal H_{B,q}$ is the bath subspace containing exactly $q$
occupations.  This ensemble is not Haar random on the complete
$\mathcal H_q$ and does not have Eq.~\eqref{eq:sectorlaw} as an exact
ensemble-average identity.  The time-resolved symmetry-breaking calculation
uses the still more restrictive product-state initialization
$|0\rangle_S\otimes|b_q\rangle_B$, with $|b_q\rangle_B$ a bath
computational-basis state of Hamming weight $q$.  The latter convention also
matches the Qiskit and QPU preparations described below.  For every protocol,
the reduced state $\bar{\rho}_S^{(q)}$ is obtained by tracing out the bath,
and Eqs.~\eqref{eq:M} and \eqref{eq:DeltaGrange} are evaluated from the
sector-averaged probe occupations.

\subsection{Symmetry-matched Floquet circuit}
\label{sec:circuit_methods}

The QPU circuit is a digital, symmetry-matched realization of the same
conserved-charge structure; it is not presented as a Trotter decomposition of
Eq.~\eqref{eq:fullH}.  Logical qubit $0$ is the probe and logical qubits
$1,\ldots,N$ form the bath.  The probe is initialized in $|0\rangle$.  For a
sector labeled by $q\in\mathcal{Q}_N$, exactly $q$ bath qubits are initialized
in $|1\rangle$, so the initial total occupation is $Q=q$.

One charge-preserving Floquet layer is
\begin{equation}
U_{\ell}^{(0)}
=
U_{\rm odd}(\boldsymbol{\theta}_{\ell}^{\rm odd})
R_z(\boldsymbol{\phi}_{\ell,2})
U_{\rm even}(\boldsymbol{\theta}_{\ell}^{\rm even})
R_z(\boldsymbol{\phi}_{\ell,1}),
\label{eq:floquet_layer}
\end{equation}
where $\ell\in\{1,\ldots,d\}$ is the layer index and $d$ is the Floquet
depth.  The product of single-qubit phase rotations is
\begin{equation}
R_z(\boldsymbol{\phi}_{\ell,r})
\equiv
\prod_{i=0}^{N}R_z^{(i)}(\phi_{\ell,r,i}),
\qquad
R_z^{(i)}(\phi)
\equiv
e^{-i\phi\sigma_z^{(i)}/2},
\label{eq:Rz_layer}
\end{equation}
where $r\in\{1,2\}$ distinguishes the two phase sublayers and each phase is
sampled independently as
$\phi_{\ell,r,i}\sim\mathcal{U}[-\pi,\pi]$.
Here, $\mathcal{U}[a,b]$ denotes the continuous uniform distribution on
$[a,b]$.

The even- and odd-bond exchange layers act on the disjoint bond sets
\begin{equation}
\mathcal{E}_{\rm even}
=\{(0,1),(2,3),\ldots\},
\qquad
\mathcal{E}_{\rm odd}
=\{(1,2),(3,4),\ldots\},
\label{eq:bond_sets}
\end{equation}
truncated at logical qubit $N$.  For
$s\in\{{\rm even},{\rm odd}\}$,
\begin{equation}
U_s(\boldsymbol{\theta}_{\ell}^{s})
\equiv
\prod_{(i,j)\in\mathcal{E}_s}
U_{ij}^{XY}(\theta_{\ell,ij}),
\label{eq:exchange_layer}
\end{equation}
where gates within each product act on disjoint qubit pairs and therefore
commute.  The implemented Qiskit gate is
\texttt{XXPlusYYGate}$(\theta,0)$, whose action in the ordered basis
$\{|00\rangle,|01\rangle,|10\rangle,|11\rangle\}$ is
\begin{equation}
U_{ij}^{XY}(\theta)
=
\begin{pmatrix}
1&0&0&0\\
0&\cos(\theta/2)&-i\sin(\theta/2)&0\\
0&-i\sin(\theta/2)&\cos(\theta/2)&0\\
0&0&0&1
\end{pmatrix}.
\label{eq:xy_gate}
\end{equation}
The exchange angles are sampled as
\begin{equation}
\theta_{\ell,ij}
=
\frac{\pi}{2}\left(1+0.30\,\xi_{\ell,ij}\right),
\qquad
\xi_{\ell,ij}\sim\mathcal{U}[-1,1].
\label{eq:theta_distribution}
\end{equation}
The full depth-$d$ unitary is
\begin{equation}
\mathcal{U}_d^{(0)}
\equiv
U_d^{(0)}U_{d-1}^{(0)}\cdots U_1^{(0)},
\label{eq:depth_unitary}
\end{equation}
where the rightmost layer acts first.  Every $R_z$ and $XX+YY$ gate commutes with $Q$;
therefore,
\begin{equation}
[\mathcal{U}_d^{(0)},Q]=0
\label{eq:circuit_charge}
\end{equation}
for the ideal circuit at any depth.

Controlled symmetry breaking is introduced after every charge-preserving
layer through bath-only rotations
\begin{equation}
U_x(\epsilon)
\equiv
\prod_{j=1}^{N}R_x^{(j)}(\epsilon),
\qquad
R_x^{(j)}(\epsilon)
\equiv
e^{-i\epsilon\sigma_x^{(j)}/2},
\label{eq:Rx_breaking}
\end{equation}
where $\epsilon$ is the rotation angle.  The perturbed layer is
\begin{equation}
U_{\ell}(\epsilon)=U_x(\epsilon)U_{\ell}^{(0)}.
\label{eq:perturbed_layer}
\end{equation}
Because $[U_x(\epsilon),Q]\neq0$ for $\epsilon\neq0$, this operation mixes
different charge sectors.  In the paired-$\epsilon$ protocol, the initial
bath basis state, all phases $\phi_{\ell,r,i}$, and all exchange variables
$\xi_{\ell,ij}$ are held fixed for a given pair; only $\epsilon$ is changed.
This isolates the effect of the symmetry-breaking layer from circuit-instance
variability.

For comparison with continuous-time notation, the weakly broken Hamiltonian
may be written
\begin{equation}
H_\epsilon=H_0+\epsilon W,
\qquad
[H_0,Q]=0,
\qquad
[W,Q]\neq0,
\label{eq:Heps}
\end{equation}
where $H_0$ is the symmetry-preserving Hamiltonian, $W$ is a charge-breaking
operator, and $\epsilon$ controls its strength.  A prethermal regime is
operationally identified when
\begin{equation}
\rho_S^{(\epsilon)}(t)\approx\rho_S^{(0)}(t),
\qquad
t\ll\tau_{\rm sb}(\epsilon),
\label{eq:pretherm}
\end{equation}
where $\rho_S^{(\epsilon)}(t)$ and $\rho_S^{(0)}(t)$ are the reduced probe
states generated by the perturbed and symmetry-preserving dynamics,
respectively, and $\tau_{\rm sb}(\epsilon)$ is the symmetry-breaking
timescale.  At finite circuit depth, the corresponding diagnostic is
\begin{equation}
M_N(d,\epsilon)
=
\operatorname{Var}_{q\in\mathcal{Q}_N}
\left[p_e^{(q)}(d,\epsilon)\right],
\label{eq:Mt}
\end{equation}
with an analogous depth-dependent
$\Delta_G^{\rm global}(d,\epsilon)$.

\subsection{Ideal, noisy, and QPU implementations}
\label{sec:qpu_methods}

The logical circuits were first simulated with Qiskit statevector evolution
and then transpiled to a line connectivity using the basis
$\{R_z,SX,X,CX\}$ for ideal post-transpilation verification
\cite{qiskit2019}.  A separate Aer feasibility calculation used depolarizing
error probabilities $2\times10^{-4}$ for one-qubit gates and
$8\times10^{-3}$ for two-qubit gates, together with asymmetric readout errors
$P(1|0)=0.015$ and $P(0|1)=0.020$.  This noise model was used only as a
stress test and was not treated as a calibration model of IBM Fez.

The QPU experiment was executed on IBM Fez with 4096 shots per circuit.  For
$N=4$, the five logical qubits were mapped to physical qubits
$[124,123,136,143,142]$; for $N=5$, the six logical qubits were mapped to
$[124,123,136,143,142,141]$.  The circuits were transpiled to the backend
instruction set at optimization level 2.  Dynamical decoupling and gate
twirling were not applied.  Science circuits were randomized in execution
order across sector $q$, depth $d$, circuit instance, and perturbation strength
$\epsilon$ to reduce sensitivity to temporal drift.

The main unpaired scan used depths
$d\in\{0,1,2,3,4\}$ for $N=4$ and
$d\in\{0,1,2,4,6\}$ for $N=5$.  The paired symmetry-breaking scan used
$d=3$ for $N=4$ and $d=6$ for $N=5$, with $\epsilon\in\{0,0.04,0.08,0.16,0.24,0.32\}$.
For the paired scan, 12 circuit pairs were used per sector for $N=4$ and
8 circuit pairs per sector for $N=5$.  Depth-zero circuits and no-exchange
circuits, in which all $XX+YY$ gates were omitted while the preparation and
$R_z$ layers were retained, were executed as controls.

Let
$\mathbf{b}=(b_0,b_1,\ldots,b_N)\in\{0,1\}^{N+1}$ denote a measured logical
bit string, where $b_0$ is the probe bit, and let
$|\mathbf{b}|\equiv\sum_{i=0}^{N}b_i$ be its Hamming weight.  For a probability
distribution $P_{q,d,\epsilon}(\mathbf{b})$ obtained from circuits prepared in
sector $q$, the probe occupation and charge-preservation probability are
\begin{subequations}
\begin{align}
p_e^{(q)}(d,\epsilon)
&=
\sum_{\mathbf{b}}b_0P_{q,d,\epsilon}(\mathbf{b}),
\label{eq:qpu_pe}\\
P_Q^{(q)}(d,\epsilon)
&=
\sum_{\mathbf{b}:|\mathbf{b}|=q}
P_{q,d,\epsilon}(\mathbf{b}).
\label{eq:qpu_PQ}
\end{align}
\end{subequations}
The reported sector value is the arithmetic mean over the circuit instances
used for the same $(N,q,d,\epsilon)$.  The quantity reported as $P_Q$ is the
subsequent arithmetic mean of $P_Q^{(q)}$ over the sampled sectors.  No charge
postselection is used in the main analysis; $P_Q$ is retained as an
independent diagnostic of symmetry preservation and hardware-induced leakage.

\subsection{Readout mitigation and uncertainty propagation}
\label{sec:mitigation_methods}

Full computational-basis readout-calibration circuits were measured for each
selected register.  Let $x$ and $y$ denote the prepared and measured basis
states, respectively, represented as integers in
$\{0,1,\ldots,2^{N+1}-1\}$.  The correlated assignment matrix is $A_{yx}\equiv P(y|x)$, where each column is estimated from the corresponding calibration counts.  A
local assignment model was also obtained by marginalizing the correlated
matrix to one $2\times2$ matrix per qubit and forming their tensor product.
For either assignment matrix $A$, the measured probability vector
$\mathbf{p}_{\rm meas}$ was corrected as
\begin{equation}
\widetilde{\mathbf{p}}
=A^{+}\mathbf{p}_{\rm meas},
\qquad
\mathbf{p}_{\rm mit}
=
\frac{[\widetilde{\mathbf{p}}]_+}
{\sum_y[\widetilde{p}_y]_+},
\label{eq:readout_mitigation}
\end{equation}
where $A^{+}$ is the Moore--Penrose pseudoinverse with numerical cutoff
$10^{-6}$ and $[z]_+\equiv\max(z,0)$ is applied elementwise.  Raw, locally
mitigated, and correlated-mitigated results were analyzed independently.

Central 68\% confidence intervals were obtained by a two-level bootstrap.  In
each bootstrap replicate, the calibration counts were first resampled so that
the local and correlated assignment matrices were reconstructed.  Circuit
instances were then sampled with replacement within each sector, and the
finite-shot counts of every selected circuit were multinomially resampled.
The complete analysis, including readout mitigation, sector averaging,
$M_N$, $\Delta_G^{\rm global}$, and $P_Q$, was repeated for that replicate.
The lower and upper interval limits are the 15.9th and 84.1st percentiles,
respectively.  This procedure propagates calibration uncertainty, shot noise,
and circuit-instance variability without changing any central-value
calculation.

\subsection{Disorder and detuning sensitivity tests}
\label{sec:disorder_methods}

The supplementary robustness simulations used the same paired circuit
realizations.  Static bath-frequency disorder was implemented by adding the
phase $\sigma_\omega\zeta_j$ to bath-qubit $j$ in every $R_z$ sublayer, where
$\sigma_\omega$ is the disorder strength and each $\zeta_j$ is a fixed
standard-normal random variable for that paired realization.  Probe--bath
detuning was implemented by adding a probe phase $\Delta$ per $R_z$ sublayer,
where $\Delta$ is the detuning angle.  Coupling disorder was implemented as
\begin{equation}
\theta_{\ell,ij}
\longrightarrow
\theta_{\ell,ij}\left(1+\sigma_g\eta_{ij}\right),
\label{eq:coupling_disorder}
\end{equation}
where $\sigma_g$ is the coupling-disorder strength and each $\eta_{ij}$ is a
fixed standard-normal random variable for the paired realization.  These
perturbations commute with $Q$ and therefore test finite-depth intra-sector
mixing rather than explicit symmetry breaking.

\section{Results}
\label{sec:results}

We now evaluate the probe-level bath-emergence criterion introduced above.
The analysis is organized in three steps.  First, we test the sector-memory
diagnostics in an exactly simulated charge-conserving Hamiltonian model.
Second, we construct a hardware-compatible Floquet circuit and validate that
it reproduces the same sector law before execution on a quantum processor.
Third, we implement the circuits on IBM Fez and test both the charge-preserving
case and a paired weak-symmetry-breaking perturbation.  

\subsection{Exact charge conservation produces a sector-dependent probe state}
\label{subsec:results_exact}

\begin{figure}[t]
    \centering
    \includegraphics[width=\linewidth]{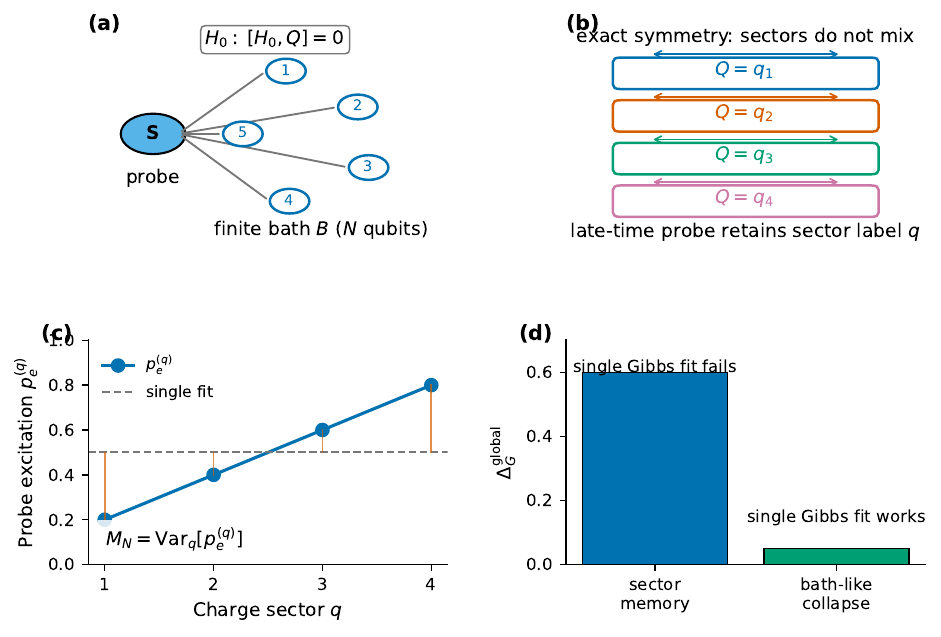}
    \caption{
    Probe-level diagnostics for finite-bath emergence.  A probe qubit \(S\)
    is coupled to a finite bath \(B\) of \(N\) qubits.  Under charge-preserving
    dynamics, initial fixed-charge sectors remain dynamically separated.
    The measured quantities are the sector-resolved probe population
    \(p_e^{(q)}\), the sector-memory variance \(M_N\), and the global
    Gibbs-fit error \(\Delta_G^{\rm global}\).  A finite environment is
    bath-like for the probe only when the sector-resolved reduced states
    collapse to a common Gibbs description.
    }
    \label{fig:concept_diagnostics}
\end{figure}

Figure~\ref{fig:concept_diagnostics} summarizes the operational structure of
the test.  In the Qiskit and IBM protocols, the probe qubit is initialized in
\(|0\rangle\), the bath is initialized in a fixed excitation sector
\(q\), and the late-time or finite-depth probe population \(p_e^{(q)}\)
is measured across sectors.  The principal exact benchmark instead uses
full-sector Haar-random pure states as specified in
Eq.~\eqref{eq:full_sector_haar_initial}; the probe-fixed exact calculation is
reported separately in the Supplemental Material (Figs. S1-S5).  If the finite
environment behaves as a single canonical bath for the probe, the values
\(p_e^{(q)}\) should collapse to a sector-independent value.  Persistent
variation of \(p_e^{(q)}\) across \(q\) gives a nonzero sector-memory
variance \(M_N\) and a nonzero global Gibbs-fit error
\(\Delta_G^{\rm global}\).

For the charge-conserving Hamiltonian \(H_0\), we evaluated the diagonal
ensemble using the full-sector Haar initialization in
Eq.~\eqref{eq:full_sector_haar_initial}.  The commutator check gives
\([H_0,Q]=0\) to numerical precision; additional Hermiticity, block-structure,
and diagonal-ensemble checks are given in Figs.~S6--S9 of the Supplemental
Material.  For the \(N=8\) bath, the sector-resolved ensemble averages follow
the uniform fixed-charge benchmark given by ~\eqref{eq:sectorlaw}. 

\begin{figure}[t]
    \centering
    \includegraphics[width=\linewidth]{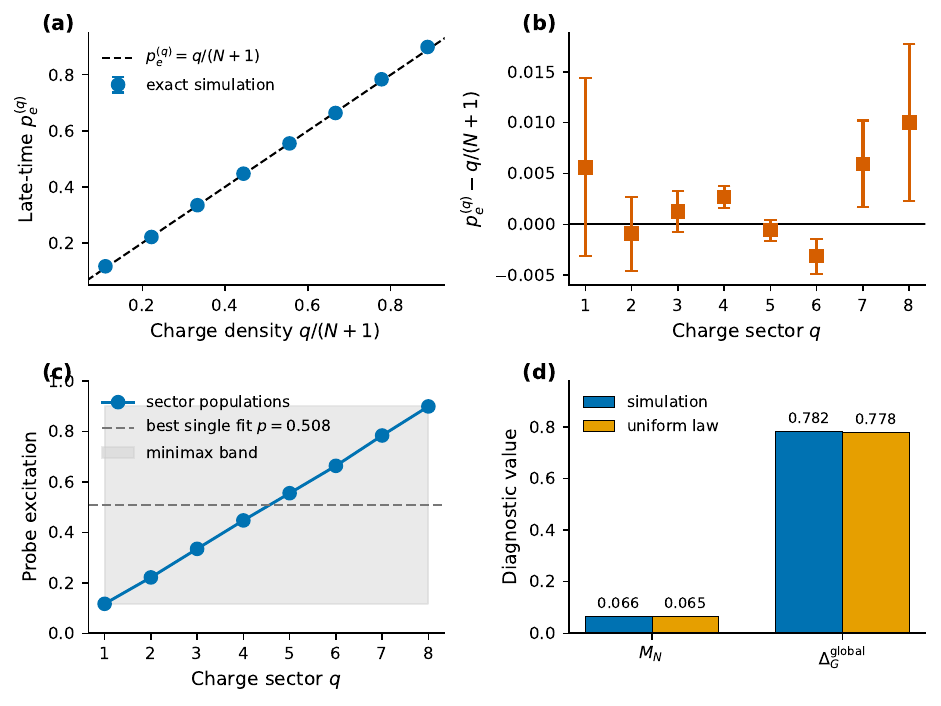}
    \caption{
    Exact Hamiltonian benchmark for sector-memory obstruction.  The
    \(N=8\) charge-conserving calculation uses Haar-random pure states over
    each complete total-charge sector; its ensemble-averaged late-time probe
    population follows \(p_e^{(q)}\simeq q/(N+1)\).  The sector-resolved probe
    states cannot be represented by one global Gibbs state: the best
    sector-independent population fit leaves a large minimax error, giving
    \(\Delta_G^{\rm global}=0.782\).
    }
    \label{fig:exact_sector_gibbs}
\end{figure}

As shown in Fig.~\ref{fig:exact_sector_gibbs}(a), the measured populations
track the line \(p_e=q/9\) over the full set \(q=1,\ldots,8\).  The largest
deviation from the sector law is approximately \(1.0\times 10^{-2}\).
The corresponding diagnostics are $M_N = 0.065$ and $\Delta_G^{\rm global}=0.782$, in close agreement with the uniform fixed-charge expectations $M_N = 0.064$ and $\Delta_G^{\rm global}=0.777$. 

The close agreement in Fig.~\ref{fig:exact_sector_gibbs} should therefore be
interpreted as a full-sector Haar typicality benchmark, not as an
initialization-independent dynamical identity.  We repeated the exact
calculation with the probe fixed in \(|0\rangle_S\) and the bath drawn from
the Haar measure on \(\mathcal H_{B,q}\), as in
Eq.~\eqref{eq:probe0_bath_haar_initial}.  In that case the sector profile does
not follow \(q/(N+1)\) exactly and becomes nonmonotonic near the upper charge
sectors (Figs. S1-S5).  Nevertheless, the \(N=8\) probe-fixed calculation gives $M_N = 0.034$ and $\Delta_G^{\rm global}=0.541$, so the sector-memory obstruction remains substantial.  These results separate the robust conclusion--persistent
sector-dependent probe states--from the initialization-dependent linear
benchmark \(q/(N+1)\).

The obstruction is not the failure of local equilibration within one sector.
Rather, it is the persistence of sector dependence across preparations.  The
best single population fit gives \(p_{\rm fit}=0.50\), corresponding to
\(|\beta_{\rm fit}|=0.031\) for \(\omega_S=1\), but it leaves a maximum
population residual of \(0.391\).  Therefore, even though the reduced
probe state is stationary within each sector, the collection
\(\{\bar{\rho}_S^{(q)}\}\) cannot be described by a single Gibbs state.
This is the finite-probe analogue of the memory retained by generalized
equilibria in the presence of conservation laws \cite{rigol2007}.

\subsection{Hardware-compatible circuit validation}
\label{subsec:results_qiskit}

The exact Hamiltonian calculation establishes the reduced-state obstruction,
but IBM hardware implements finite-depth circuits rather than diagonal
ensembles.  We therefore constructed a hardware-compatible Floquet circuit
with the same symmetry structure, as discussed in the ~\ref{sec:circuit_methods}.  

\begin{figure}[t]
    \centering
    \includegraphics[width=\linewidth]{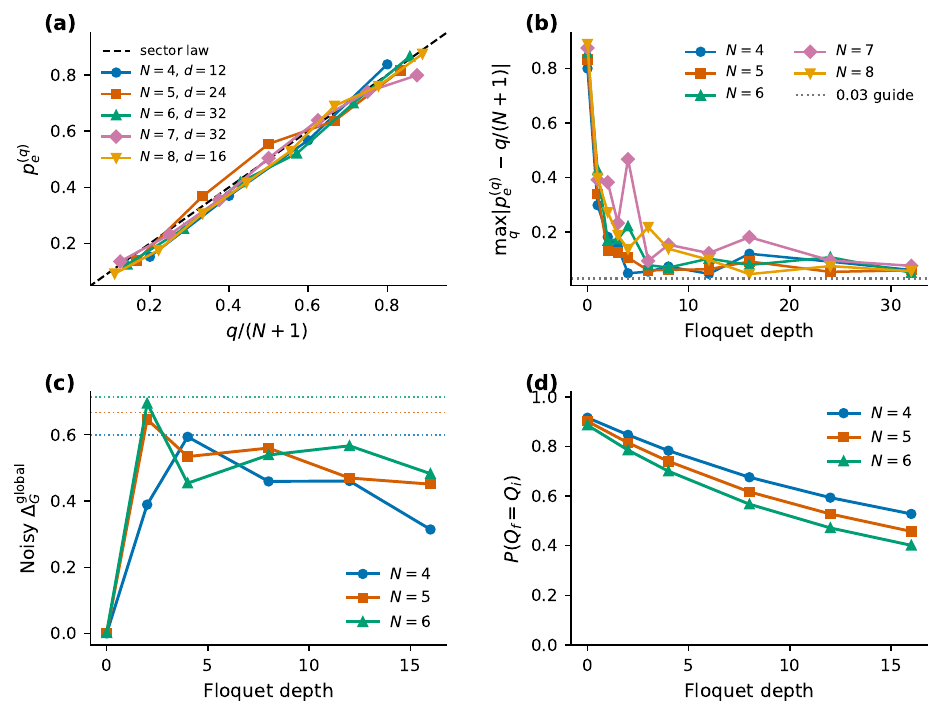}
    \caption{
    Hardware-compatible circuit validation.  Ideal and transpiled circuit
    simulations reproduce the sector law \(p_e^{(q)}\simeq q/(N+1)\).
    Simple noisy simulations identify \(N=4\) as the most robust hardware
    target: at depth \(d=4\), the noisy circuit gives
    \(\Delta_G^{\rm global}=0.593\), close to the ideal value \(0.60\),
    while retaining a visible charge-preservation signal.
    }
    \label{fig:qiskit_validation}
\end{figure}

Figure~\ref{fig:qiskit_validation} shows the ideal and noisy circuit
validation; complete depth sweeps are provided in Figs.~S10 and S11 of the
Supplemental Material.  The ideal circuits were simulated before and after transpilation
to a line-connected native-gate model.  Transpilation did not change the
sector-resolved probabilities within numerical precision, and the ideal
transpiled circuits retained exact charge conservation.  For
\(N=4,\ldots,8\), the best-depth ideal circuits produced monotonic
sector ordering and followed the expected sector law.  For example, the
ideal transpiled \(N=8\) circuit gave $M_N=0.069$ and $\Delta_G^{\rm global}=0.782$, close to the corresponding uniform-sector values
\(0.064\) and \(0.777\).

The noisy simulations determine which system sizes are experimentally useful.
For \(N=4\), the depth-\(d=4\) noisy circuit gave $M_N=0.048$ and $ \Delta_G^{\rm global}=0.593$.  The corresponding ideal
values are \(M_N=0.05\) and \(\Delta_G^{\rm global}=0.6\).
For \(N=5\) and \(N=6\), the sector ordering remained visible, but the noisy
charge-preservation probability decreased.  This motivated the hardware
choice: \(N=4\) as the primary target and \(N=5\) as a secondary robustness
test.  This circuit-level validation is distinct from recent superconducting
processor studies of thermalization, dynamic thermalization, and bath-emulated
transport \cite{zhu2022,perrin2025,zhang2024}; here the observable is the
ordinary reduced state of one local probe.

To distinguish the deliberately broad sector scan from a physically narrow preparation, we also evaluated binomially weighted sector ensembles ($w_q(\nu)=\binom{N}{q}\nu^q(1-\nu)^{N-q}$). In the fixed-sector benchmark (~\eqref{eq:sectorlaw}), the weighted sector-memory variance obeys
\begin{equation}
M_N[w]\simeq \frac{\mathrm{Var}_w(q)}{(N+1)^2}.
\end{equation}
For a binomial sector distribution centered at fixed excitation density ($\nu$), this gives
\begin{equation}
M_N^{\rm binomial}\simeq
\frac{N\nu(1-\nu)}{(N+1)^2}=O(N^{-1}).
\end{equation}
The strict full-support minimax spread remains (O(1)), because the binomial distribution has exponentially small tails at $q=0, N$. However, for a central probability window containing fixed mass ($\alpha$), the sector-law prediction is
\begin{equation}
\Delta_{G,\alpha}^{\rm global}\simeq
\frac{q_+(\alpha)-q_-(\alpha)}{N+1}
=O(N^{-1/2}).
\end{equation}
Qiskit simulations of the same charge-preserving Floquet circuits confirm these scalings up to finite-depth mixing errors and, in noisy simulations, charge-leakage-induced compression of the sector contrast (see Fig. S11).

\subsection{IBM Fez observation of sector memory}
\label{subsec:results_ibm_sector}

The IBM Fez experiment used \(4096\) shots per circuit and randomized circuit
order across sector, depth, initial state, and perturbation strength.  For
\(N=4\), the physical path was
\([124,123,136,143,142]\); for \(N=5\), the path was
\([124,123,136,143,142,141]\).  Full readout calibration circuits were
collected for the selected qubits.  The correlated readout assignment
matrices were well-conditioned, with condition numbers \(1.087\) for the
five-qubit register and \(1.103\) for the six-qubit register.  Local and
correlated readout mitigation gave indistinguishable conclusions; the backend
layout, calibration matrices, and control data are summarized in Figs.~S12--S14
of the Supplemental Material.

Figure~\ref{fig:ibm_sector_memory} shows the main hardware observation.
For the primary \(N=4\) experiment, the cleanest charge-preserving result
occurs at depth \(d=3\) and \(\epsilon=0\).  After local readout mitigation,
the sector-resolved populations were
\begin{equation}
\begin{array}{c|cccc}
q & 1 & 2 & 3 & 4 \\
\hline
p_e^{(q)} & 0.170 & 0.343 & 0.544 & 0.728
\end{array}
\end{equation}
to be compared with the ideal values \(q/5=\{0.2,0.4,0.6,0.8\}\).
The corresponding diagnostics were $M_N=0.044$ and $\Delta_G^{\rm global}=0.558$ with mean charge preservation \(0.902\).  The raw data already showed
the same sector ordering, with
\(\Delta_G^{\rm global}=0.552\).

\begin{figure}[t]
    \centering
    \includegraphics[width=\linewidth]{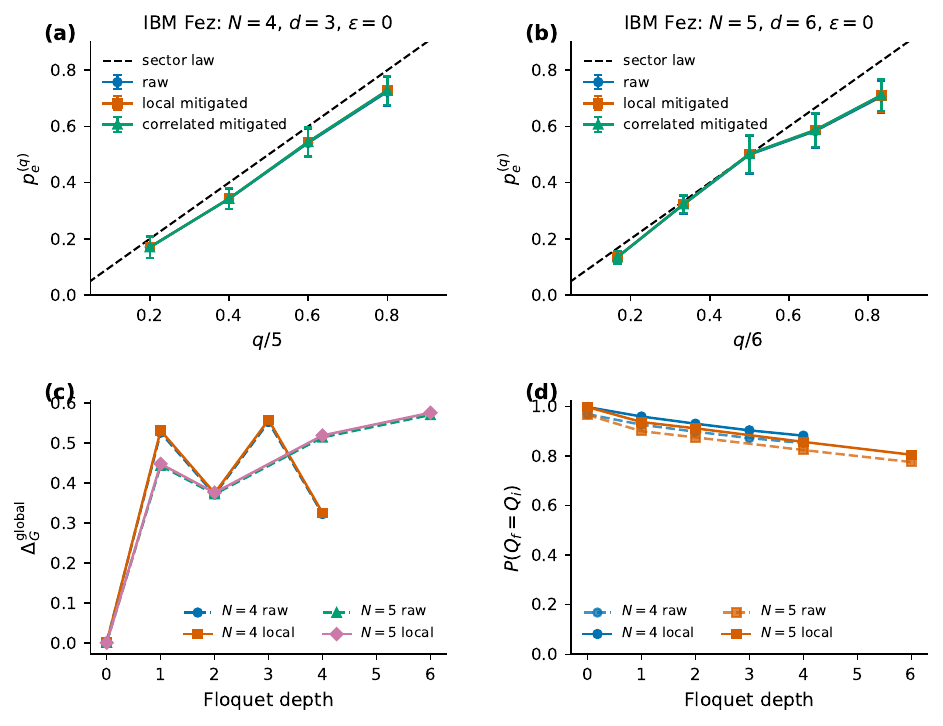}
    \caption{
    IBM Fez observation of probe-level sector memory.  The \(N=4\),
    \(d=3\), \(\epsilon=0\) data show clear monotonic sector ordering and
    a large nonzero \(\Delta_G^{\rm global}\).  The \(N=5\), \(d=6\),
    \(\epsilon=0\) data show that the effect survives at a larger bath size,
    although with stronger compression due to circuit depth and hardware
    noise.  Raw and readout-mitigated analyses give the same qualitative
    conclusion.
    }
    \label{fig:ibm_sector_memory}
\end{figure}

The \(N=5\) charge-preserving experiment at \(d=6\) also retained a clear
sector dependence.  The local readout-mitigated populations were
\begin{equation}
\begin{array}{c|ccccc}
q & 1 & 2 & 3 & 4 & 5 \\
\hline
p_e^{(q)}
& 0.134 & 0.322 & 0.502 & 0.587 & 0.710 ,
\end{array}
\end{equation}
compared with \(q/6\).  The resulting diagnostics were $M_N=0.041$ and $\Delta_G^{\rm global}=0.575$, with mean charge preservation \(0.804\).  Although the \(N=5\) data are
more compressed than the ideal sector law, they remain incompatible with a
single sector-independent Gibbs population.

Several controls rule out a trivial readout or preparation origin of the
observed sector dependence.  At depth zero, the probe remains near
\(|0\rangle\), as expected.  In no-exchange controls, where the same initial
states and phase layers are used but all \(XX+YY\) exchange gates are
removed, the maximum raw probe excitation is \(2.4\times 10^{-3}\) for
\(N=4\) and \(3.4\times 10^{-3}\) for \(N=5\).  After local readout
mitigation, the corresponding maxima are \(3.4\times 10^{-4}\) and
\(1.44\times 10^{-3}\).  Thus the finite-depth sector dependence in
Fig.~\ref{fig:ibm_sector_memory} is generated by the exchange dynamics, not
by static readout bias.

A two-level bootstrap uncertainty analysis was performed on the hardware data. In each bootstrap replicate, the readout-calibration counts for every prepared computational-basis calibration state are first resampled, and both the correlated assignment matrix and the corresponding local tensor-product assignment model are reconstructed. The data are then resampled by drawing circuit instances with replacement within each charge sector and applying multinomial shot resampling to the measured counts of each selected circuit. Raw, locally readout-mitigated, and correlated-readout-mitigated sector populations are recomputed from the same bootstrap replicate, and the derived quantities $M_N$, $\Delta_G^{\rm global}$, and $P_Q=P(Q_f=Q_i)$ are recalculated. The reported intervals are central 68\% bootstrap intervals, i.e., the 15.9th and 84.1st percentiles.

This procedure propagates finite-shot uncertainty in the actual "science" circuits, circuit-instance variability within each charge sector, and finite-shot uncertainty in the readout-calibration matrices. For the primary IBM Fez $N=4,d=3,\epsilon=0$ data, the locally mitigated analysis gives $M_N=0.044^{+0.011}_{-0.008}$ and $\Delta_G^{\rm global}=0.558^{+0.062}_{-0.057}$. For the $N=5,d=6,\epsilon=0$ data, the locally mitigated analysis gives $M_N=0.041^{+0.009}_{-0.006}$ and $\Delta_G^{\rm global}=0.576^{+0.063}_{-0.052}$. The raw, local-mitigated, and correlated-mitigated estimates agree within these intervals, confirming that the observed sector-memory obstruction is not a readout-mitigation artifact.

In these experiments, the reduced probe state is diagonal in the probe-number basis due to a selection rule. A fixed total-charge state has support only on $|0\rangle_S\otimes\mathcal{H}_{B,q}$ and $|1\rangle_S\otimes\mathcal{H}_{B,q-1}$. The reduced off-diagonal element $\langle0|\rho_S|1\rangle$ would require the same bath state to appear in both branches, but those branches lie in orthogonal bath-charge sectors. Thus, for fixed-sector charge-preserving dynamics, the probe coherence vanishes exactly. We did not perform a separate Ramsey or tomographic hardware measurement. Thus, the reported Gibbs obstruction does not rely on tomography: the measured population spread alone rules out a single common Gibbs state across sectors.

\begin{table*}[t]
\caption{
Summary of the principal diagnostics.  The exact-Hamiltonian result is obtained
from the diagonal ensemble, the Qiskit results validate the hardware-compatible
circuits, and the IBM Fez results correspond to measured hardware data. Here \(n_q\) is the number of physical qubits.  For the main unpaired IBM Fez
data, the tabulated values are local-readout-mitigated values.  For the
paired-\(\epsilon\) scan, the tabulated values are raw paired values; the
corresponding readout-mitigated trends are reported in the Supplement.
}
\label{tab:main_diagnostics}
\centering
\small
\setlength{\tabcolsep}{4.5pt}
\begin{ruledtabular}
\begin{tabular}{lccccclcccc}
Platform & \(N\) & \(n_q\) & depth & \(\epsilon\) & data
& analysis & \(M_N\) & \(\Delta_G\) & \(P_Q\) & residual \\
\hline
Exact Hamiltonian (full-sector Haar) & 8 & 9 & -- & 0 & exact & diag. ens. & 0.0655 & 0.782 & 1.00  & 0.391 \\
Qiskit ideal           & 4 & 5 & 12 & 0     & sim.  & ideal      & 0.0636 & 0.684 & 1.00  & 0.0471 \\
Qiskit noisy           & 4 & 5 & 4  & 0     & sim.  & noisy      & 0.0484 & 0.594 & 0.783 & 0.0196 \\
IBM Fez                & 4 & 5 & 3  & 0     & hw.   & mitigated  & 0.0440 & 0.558 & 0.903 & 0.0719 \\
IBM Fez                & 5 & 6 & 6  & 0     & hw.   & mitigated  & 0.0410 & 0.576 & 0.805 & 0.123 \\
IBM Fez paired         & 4 & 5 & 3  & 0     & hw.   & raw        & 0.0427 & 0.553 & 0.847 & 0.0762 \\
IBM Fez paired         & 4 & 5 & 3  & 0.320 & hw.   & raw        & 0.0386 & 0.529 & 0.668 & 0.0888 \\
IBM Fez paired         & 5 & 6 & 6  & 0     & hw.   & raw        & 0.0296 & 0.475 & 0.743 & 0.138 \\
IBM Fez paired         & 5 & 6 & 6  & 0.320 & hw.   & raw        & 0.0232 & 0.416 & 0.450 & 0.145 \\
\end{tabular}
\end{ruledtabular}
\end{table*}

\subsection{Paired symmetry-breaking scan}
\label{subsec:results_paired_epsilon}

We next tested whether explicit symmetry breaking reduces the sector-memory
diagnostics.  The perturbation was implemented by applying \(R_x(\epsilon)\)
rotations to the bath qubits after each charge-preserving Floquet layer.
Because weakly broken conservation laws are expected to generate long-lived
prethermal regimes before full thermalization \cite{mallayya2019}, the
hardware question is not whether \(M_N\) immediately vanishes, but whether
it decreases as charge conservation is progressively violated.

To make the \(\epsilon\)-dependence causal, the hardware scan was paired:
for each \((N,q,d,\mathrm{pair})\), the initial bath state, random \(R_z\)
phases, and \(XX+YY\) exchange angles were held fixed, and only
\(\epsilon\) was changed.  The scan used $\epsilon=0,\;0.04,\;0.08,\;0.16,\;0.24,\;0.32$. 
For \(N=4\), the depth was \(d=3\) with 12 paired instances per sector.
For \(N=5\), the depth was \(d=6\) with 8 paired instances per sector.

\begin{figure}[t]
    \centering
    \includegraphics[width=\linewidth]{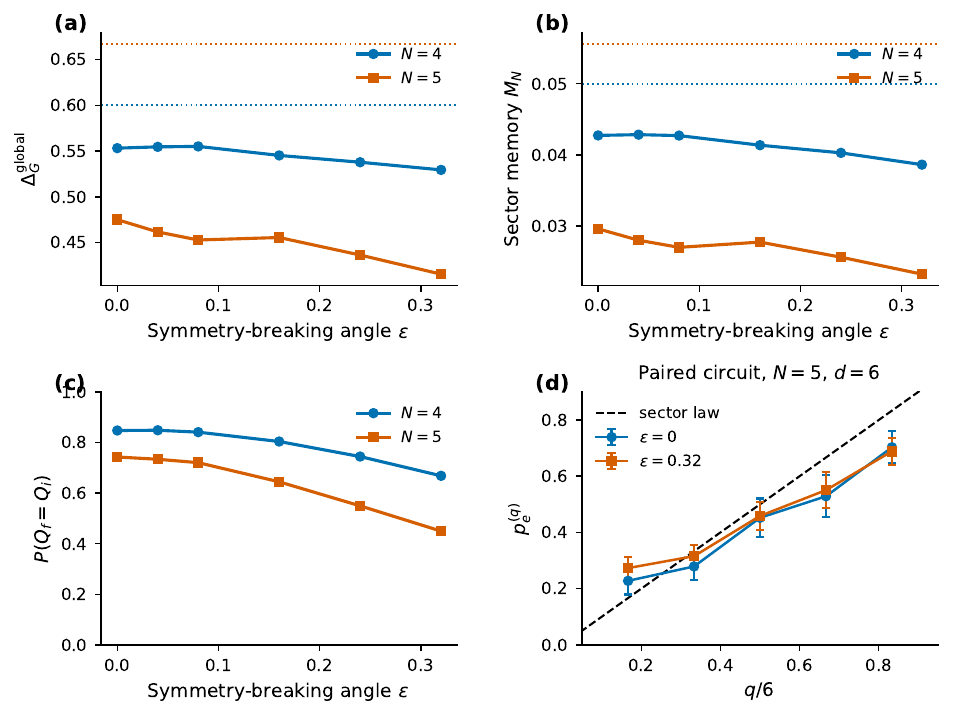}
    \caption{
    Paired symmetry-breaking scan on IBM Fez.  For each paired circuit,
    the initial state and charge-preserving Floquet layers are identical
    across \(\epsilon\); only the bath \(R_x(\epsilon)\) rotations are
    changed.  Increasing \(\epsilon\) reduces \(M_N\) and
    \(\Delta_G^{\rm global}\) and decreases charge preservation, but the
    sector ordering remains visible for both \(N=4\) and \(N=5\).
    }
    \label{fig:paired_epsilon}
\end{figure}

As shown in Figure~\ref{fig:paired_epsilon},  complete hardware
depth--\(\epsilon\) maps and pair-resolved changes are shown in Figs.~S15
and S16 of the Supplemental Material.  For \(N=4\),
increasing \(\epsilon\) from 0 to 0.32 reduced
\begin{equation}
    M_N:\;0.042\rightarrow 0.038,
    \qquad
    \Delta_G^{\rm global}:\;0.553\rightarrow 0.529.
\end{equation}
Over the same interval, the charge-preservation probability decreased from
\(0.847\) to \(0.668\).  Thus the perturbation measurably breaks charge
conservation, but it does not erase the sector structure.

The suppression is slightly stronger for \(N=5\).  From \(\epsilon=0\) to
\(\epsilon=0.32\),
\begin{equation}
    M_N:\;0.029\rightarrow 0.023,
    \qquad
    \Delta_G^{\rm global}:\;0.475\rightarrow 0.415,
\end{equation}
while charge preservation decreases from \(0.742\) to \(0.449\).
At all measured \(\epsilon\), however, the sector ordering remains monotonic.
Therefore the paired scan shows the onset of symmetry-breaking suppression
of sector memory, not a complete crossover to a sector-independent Gibbs
state.  This distinction is important: the finite hardware circuits remain
in a constrained, prethermal-like regime on the accessible depth scale.

To determine whether a sector-independent description emerges at larger depth, we supplemented the fixed-depth hardware scan with a depth-resolved paired-($\epsilon$) Qiskit study. For each $(N,q,\mathrm{pair})$, the initial bath state and all charge-preserving Floquet angles were held fixed as depth was increased; across $\epsilon$, only the bath $R_x(\epsilon)$ rotations were changed. We then evaluated $M_N(d,\epsilon)$, $\Delta_G^{\rm global}(d,\epsilon)$, and $P(Q_f=Q_i;d,\epsilon)$ in ideal statevector simulations and in noisy transpiled Aer simulations. The simulations show that the hardware range ($\epsilon\leq 0.32$) lies in the onset regime: charge preservation is strongly reduced, and $M_N$ and $\Delta_G^{\rm global}$ decrease with depth, but a sector-independent threshold is not reached by the largest simulated depths. Using the operational criterion $M_N\leq 0.005$ and $\Delta_G^{\rm global}\leq 0.10$, all $N=4,5,6$ simulations with $\epsilon\leq 0.32$ give $d_{\rm sb}>32$ in the ideal data and $d_{\rm sb}>24$ in the noisy data. For stronger perturbations, the crossover becomes visible: for example, at $\epsilon=0.50$, the threshold is reached at $d_{\rm sb}=16,24,32$ in the ideal $N=4,5,6$ simulations and at $d_{\rm sb}=16,24,24$ in the corresponding noisy simulations. Thus, the measured hardware scan is consistent with the early-depth part of a symmetry-breaking crossover, while the fully sector-independent regime requires larger $\epsilon$, larger depth, or both.  The complete depth-resolved scans, half-decay and threshold depths, terminal compression, and sector populations are reported in Figs.~S17--S22 and Table~S2 of the Supplemental Material.

We also tested whether the observed sector-memory obstruction is sensitive to frequency disorder, coupling inhomogeneity, or probe--bath detuning. In the circuit representation, bath-frequency disorder and probe detuning enter as additional number-conserving $R_z$ phases, while coupling disorder enters as variations of the $XX+YY$ exchange angles. These perturbations preserve total excitation number at $\epsilon=0$, and therefore do not remove the fixed-charge-sector structure. Their effect is instead to modify the finite-depth rate at which the circuit explores each charge sector.

The resulting ideal and noisy Qiskit simulations show that the sector-memory signal is robust to moderate disorder; representative values are collected in Table~S3 of the Supplemental Material. At the target depths $d=3,6,8$ for $N=4,5,6$, respectively, bath-frequency disorder, exchange-angle disorder, and probe--bath detuning change the finite-depth residual from the uniform-sector benchmark
but do not generically collapse the sector-resolved probe populations to a common value. For example, in the ideal $N=5,d=6$ circuit, the clean value $\Delta_G^{\rm global}=0.656$ remains sizable under strong isolated perturbations: $\Delta_G^{\rm global}=0.568$ for bath-frequency disorder strength $\sigma_\omega=1$, $0.594$ for probe detuning $\Delta=2$, and $0.529$ for coupling-disorder strength $\sigma_g=0.75$. The corresponding noisy transpiled simulations give the same qualitative conclusion, with $\Delta_G^{\rm global}=0.523$, $0.524$, and $0.455$, respectively, compared with the clean noisy value $0.552$. Thus, even when the sector law is distorted by finite-depth mixing errors and hardware-like noise, the Gibbs-fit obstruction remains nonzero.

The largest finite-depth effect is produced by strong probe--bath detuning or combined disorder, which can suppress excitation exchange, compress the measured sector contrast, and increase the residual from $q/(N+1)$. This behavior should not be interpreted as emergence of a sector-independent Gibbs state: the perturbations are number conserving, so the ideal charge preservation remains unity, and the observed compression reflects slower or less complete intra-sector mixing at the accessible depth. In the noisy simulations, the charge-preservation probability is instead set mainly by the hardware-like noise model and is nearly insensitive to the disorder strength. Overall, disorder and detuning affect the visibility and convergence rate of the finite-depth sector law, but they do not remove the underlying sector-memory obstruction whenever the dynamics remains charge preserving.

The exact simulation, hardware-compatible circuit validation, and IBM Fez experiments give a consistent picture.  The full-sector Haar exact benchmark follows \(p_e^{(q)}\simeq q/(N+1)\), whereas the probe-fixed exact calculation does not obey this relation exactly.  Both initializations nevertheless produce probe populations controlled by the conserved sector \(q\).  The resulting nonzero
\(M_N\) and \(\Delta_G^{\rm global}\) exclude a single canonical Gibbs
description across sectors.  The same diagnostic survives in finite-depth hardware circuits on IBM Fez for \(N=4\) and \(N=5\).  Weak symmetry breaking reduces the sector-memory diagnostics and increases charge leakage, but does not eliminate sector dependence over the measured depths. These results support the central claim: a finite programmable qubit
environment can equilibrate a local probe within each constrained sector
while still failing to act as a sector-independent canonical bath.  Bath
emergence at the probe level therefore requires not only local relaxation, but also the loss of preparation-dependent sector memory.

\section{Conclusion}
\label{sec:conclusion}

We have studied bath emergence from the perspective of a single probe qubit
coupled to a finite programmable environment.  The central point is that local
relaxation of the probe is not, by itself, sufficient to establish a canonical
bath description.  When the dynamics conserves total excitation number, the
initial charge sector remains encoded in the reduced probe state.  This
produces sector-resolved probe populations \(p_e^{(q)}\), a nonzero
sector-memory variance \(M_N\), and a nonzero global Gibbs-fit error
\(\Delta_G^{\rm global}\).  The resulting obstruction is operational: if the
same finite environment gives different stationary probe populations for
different charge-sector preparations, then no single inverse temperature
\(\beta\) can describe all reduced probe states.

Exact Hamiltonian simulations confirm this mechanism.  For a charge-conserving
\(N=8\) bath, the full-sector Haar ensemble used in the principal benchmark
follows \(p_e^{(q)}\simeq q/(N+1)\).  The observed values
\(M_N=0.065\) and \(\Delta_G^{\rm global}=0.782\) are close to the
maximally mixed-sector predictions \(0.064\) and \(0.777\), respectively.
This linear relation is not universal: when the probe is fixed initially in
\(|0\rangle_S\) and the bath alone is Haar random at fixed charge, the
sector profile deviates from \(q/(N+1)\), while remaining strongly
sector dependent with \(M_N=0.034\) and
\(\Delta_G^{\rm global}=0.541\).  Thus the robust conclusion is the
failure of one global Gibbs state across sectors, not exact universality of the
linear benchmark.

We then translated the diagnostic into a hardware-compatible circuit model.
The ideal Floquet circuit uses only excitation-preserving operations:
random \(R_z\) phases and \(XX+YY\) exchange gates.  Qiskit simulations show
that these circuits reproduce the same sector-memory structure after
transpilation.  Noisy simulations identify \(N=4\) as the most robust hardware
target, with a large predicted \(\Delta_G^{\rm global}\) and measurable charge
preservation at hardware-accessible depths.

The IBM Fez experiments observe the predicted sector dependence directly.  In
the primary \(N=4\), \(d=3\), \(\epsilon=0\) run, local readout mitigation gives
\(M_N = 0.044, \Delta_G^{\rm global}=0.558, P(Q_{\rm final}=Q_{\rm initial})=0.902\).
The \(N=5\), \(d=6\), \(\epsilon=0\) run also retains a sizable sector
dependence, with \(M_N=0.041\) and
\(\Delta_G^{\rm global}=0.575\).  Depth-zero and no-exchange controls show
that the observed probe excitation is generated by the exchange dynamics rather
than by preparation or readout bias.  Raw, locally mitigated, and correlated
readout-mitigated analyses give the same qualitative conclusion.

Finally, we tested controlled symmetry breaking using paired hardware circuits.
For each pair, the initial state and charge-preserving Floquet layers were
held fixed, while only the bath \(R_x(\epsilon)\) perturbation was varied.
Increasing \(\epsilon\) decreases charge preservation and modestly suppresses
\(M_N\) and \(\Delta_G^{\rm global}\).  For \(N=5\), for example,
\(\Delta_G^{\rm global}\) decreases from approximately \(0.475\) at
\(\epsilon=0\) to \(0.415\) at \(\epsilon=0.32\), while charge preservation
drops from approximately \(0.743\) to \(0.450\).  The sector ordering,
however, remains visible over the measured depth range.  The hardware data
therefore show the onset of symmetry-breaking suppression of sector memory,
not complete crossover to a sector-independent Gibbs state.

Overall, these results establish a finite-size, experimentally accessible
criterion for probe-level bath emergence.  A finite programmable environment
may produce stationary reduced probe states within each constrained sector
while still failing to act as a canonical bath across preparations.  In this
sense, bath emergence requires not only local equilibration, but also the loss
of conserved-sector memory.  The combination of exact simulation,
hardware-compatible circuit validation, and IBM Fez measurements provides a
direct route for testing this distinction on present-day quantum processors.

\section*{Data Availability}

The data supporting the findings of this study, together with the numerical-analysis scripts, Qiskit simulation notebooks, IBM Quantum hardware outputs, circuit metadata, raw measurement counts, readout-calibration data, processed observables, and figure-generation routines, are publicly available in the GitHub repository at \url{https://github.com/ramkrip/QTD}. The repository also contains the paired-$\epsilon$ symmetry-breaking data, finite-size-scaling analyses, disorder- and detuning-sensitivity tests, and the information required to reproduce the principal figures and tables. IBM Quantum hardware results may not be reproduced bit-for-bit because device calibrations and operating conditions vary over time; however, the circuit definitions, physical-qubit layouts, execution metadata, raw counts, and analysis procedures used in this work are provided.

\bibliography{references}
\bibliographystyle{apsrev4-2}

\end{document}